# Deep learning-based survival prediction for multiple cancer types using histopathology images

Ellery Wulczyn[1,†], David F. Steiner[1,†], Zhaoyang Xu[1], Apaar Sadhwani[1], Hongwu Wang[1], Isabelle Flament[2], Craig H. Mermel[1], Po-Hsuan Cameron Chen[1], Yun Liu[1,‡,*], Martin C. Stumpe[3,‡]

[1]Google Health, Palo Alto, CA, USA.
[2]Work done at Google Health via Advanced Clinical
[3]Work done at Google Health. Present address: Tempus Labs Inc, Chicago, United States
[†]Equal contribution
[‡]Equal contribution
*liuyun@google.com

# Abstract

**Background**

Providing prognostic information at the time of cancer diagnosis has important implications for treatment and monitoring. Although cancer staging, histopathological assessment, molecular features, and clinical variables can provide useful prognostic insights, improving risk stratification remains an active research area.

**Methods and findings**

We developed a deep learning system (DLS) to predict disease specific survival across 10 cancer types from The Cancer Genome Atlas (TCGA). We used a weakly-supervised approach without pixel-level annotations, and tested three different survival loss functions. The DLS was developed using 9,086 slides from 3,664 cases and evaluated using 3,009 slides from 1,216 cases. In multivariable Cox regression analysis of the combined cohort including all 10 cancers, the DLS was significantly associated with disease specific survival (hazard ratio of 1.58, 95% CI 1.28-1.70, p<0.0001) after adjusting for cancer type, stage, age, and sex. In a per-cancer adjusted subanalysis, the DLS remained a significant predictor of survival in 5 of 10 cancer types. Compared to a baseline model including stage, age, and sex, the c-index of the model demonstrated an absolute 3.7% improvement (95% CI 1.0-6.5) in the combined cohort. Additionally, our models stratified patients within individual cancer stages, particularly stage II (p=0.025) and stage III (p<0.001).

**Conclusions**

By developing and evaluating prognostic models across multiple cancer types, this work represents one of the most comprehensive studies exploring the direct prediction of clinical outcomes using deep learning and histopathology images. Our analysis demonstrates the potential for this approach to provide significant prognostic information in multiple cancer types, and even within specific pathologic stages. However, given the relatively small number of cases and observed clinical events for a deep learning task of this type, we observed wide confidence intervals for model performance, thus highlighting that future work will benefit from larger datasets assembled for the purposes for survival modeling.

# Introduction

The ability to provide prognostic information in oncology can significantly impact clinical management decisions such as treatment and monitoring. One of the most common systems for this purpose is the American Joint Committee on Cancer (AJCC) "TNM" cancer staging system, whereby tumors are classified by primary tumor size/extent (T), lymph node involvement (N), and the presence or absence of distant metastasis (M). Although TNM staging is useful and well-studied, there is room for improvement in some settings, with ongoing efforts to develop improved prediction strategies that incorporate information such as clinical variables[1,2], genetic information[3,4], and histomorphological features including tumor grade[5]. In this regard, computational image analysis of tumor histopathology offers an emerging approach to further improve patient outcome predictions by learning complex and potentially novel tumor features associated with patient survival.

In recent years, deep learning has been shown to recognize objects[6] and diagnose diseases from medical images[7,8] with impressive accuracy. In pathology, prior studies have reported deep learning models with performance on par with human experts for diagnostic tasks such as tumor detection and histologic grading[8–10]. These models are typically trained on millions of small image patches taken from whole-slide images (WSIs) of digitized pathology slides that have had specific features of interest labeled by pathologists, often involving detailed, hand-drawn annotations. The reliance on expert annotation has two notable disadvantages. Firstly, these annotations are laborious for experts, requiring hundreds to thousands of hours per prediction task of interest and limiting the ability to quickly extend to new applications such as other cancer types or histologic features. Secondly, the annotations explicitly enforce that the learned morphologic features are correlated with the known patterns being annotated.

By contrast, a different line of work focuses on directly learning morphologic features associated with survival without reliance on expert annotation for known pathologic features or regions of interest. Such approaches instead provide the machine learning models with a single "global" label per slide or case, such as a specimen's mutational status or a patient's clinical outcome. The task of predicting clinical outcomes from WSIs is particularly challenging due to the large size of these images (approximately 100,000×100,000 pixels at full resolution) along with the notion that the morphologic features associated with survival may, in principle, appear in any part of the imaged tissue. The large amount of image data in conjunction with morphological heterogeneity and unknown discriminative patterns result in an especially challenging, "weakly-supervised" learning problem.

Several prior efforts using machine learning and WSIs to address the survival prediction problem have used data from The Cancer Genome Atlas (TCGA), the largest publicly available database to our knowledge of digitized WSIs paired with clinical and molecular information[11–17]. These prior works have used feature-engineering approaches[13,16], leveraged annotated regions of interest[12,18,19], focused on learning of known histologic features[17] and/or developed models to directly predict survival for an individual cancer type. Here, we build on and extend prior work by developing an end-to-end deep learning system (DLS) to directly predict patient survival in multiple cancer types, training on whole-slide histopathology images without leveraging expert annotations or known features of interest. We test several loss functions to address the problem of right-censored patient outcomes, a convolutional neural network that is directly optimized to extract prognostic features from raw image data, and an image subsampling method to tackle the large image problem.

We evaluated our DLS's ability to improve risk stratification relative to the baseline information of TNM stage, age, and sex for 10 cancer types from TCGA. Though we observed improved risk stratification based on the model predictions for several cancer types, effect sizes were difficult to estimate precisely due to the limited number of cases and clinical events present in TCGA (350-1000 cases and 60-300 events per cancer type). While the results reported here provide support for the feasibility of developing weakly supervised deep learning models to predict patient prognosis from whole-slide images across multiple cancer types, future work exploring and validating the potential of deep learning applications for this task will require larger, clinically representative datasets.

## Methods

### Data
Digitized whole-slide images of hematoxylin-and-eosin- (H&E-) stained specimens were obtained from TCGA[20] and accessed via the Genomic Data Commons Data Portal (https://gdc.cancer.gov). Images from both diagnostic formalin-fixed paraffin-embedded (FFPE) slides and frozen specimens were included. Based on initial experiments as well as differences in the proportion of available FFPE images across cancer types (i.e. TCGA studies), both the FFPE and frozen WSIs available for each patient were used for training and case-level predictions. Each case contained 1-10 slides (median: 2). Clinical data (including approximated disease specific survival) were obtained from the TCGA Pan-Cancer Clinical Data Resource[21] and the Genomic Data Commons.

Of the TCGA studies for which cancer stage data were available, we chose the 10 studies with the highest number of cases and survival events. Clinical stage was used only for ovarian serous cystadenocarcinoma (OV), which did not have pathologic stage data available but was included given the high number of observed events. Cutaneous melanoma (SKCM) was excluded as it was not restricted to primary, untreated tumors[14,22]. Thyroid carcinoma (THCA) was excluded because only 14 of 479 cases had an observed event. Cases with missing data for either pathologic stage, age, sex, or disease specific survival were excluded from evaluation, whereas only cases missing disease specific survival were excluded from model development (training and tuning).

For each TCGA study, cases were split into train, tune, and test sets in a 2:1:1 ratio. To ensure representative splits given the small number of cases, split assignment was further stratified on whether the time of disease specific survival event was observed, and the time-to-event (discretized into 3 intervals based on the $25^{th}$ and $75^{th}$ percentiles). Across all cancer types,

4,880 cases (12,095 images) where used for training and tuning. The remaining 1,216 cases (3,009 images) where used for evaluation (Table 1).

**Deep Learning System (DLS)**

*Neural Network Architecture*

The core element of our deep learning system (DLS) consisted of multiple convolutional neural network (CNN) modules with shared weights, and an average pooling layer that merges image features computed by these modules (Figure 1). Our CNN consisted of layers of depth-wise separable convolution layers, similar to the MobileNet[23] CNN architecture. The layer sizes and the number of layers were tuned for each study via a random grid-search (see Supplementary Table S2). We chose this family of architectures because they contain relatively few parameters compared to other modern CNN architectures, which speeds up training and helps to reduce the risk of overfitting. Each CNN module took as input a randomly selected image patch from the slides in each case, such that when multiple patches were sampled, probabilistically at least one patch was likely to be informative of the outcome. Specifically, if the frequency of informative patches on a slide is $p$, the probability of not sampling any informative patch in $n$ patches decays exponentially with $n$: $(1-p)^n$, shrinking towards zero with even moderate values of $n$. This approach thus handles the weak label nature of survival prediction on large images, where the location of the informative region in the image or set of images is unknown. Furthermore, this approach naturally generalizes to multiple slides per case. During each training iteration, the n patches were sampled randomly, further ensuring that informative patches were sampled across training iterations.

Each patch was of size 256 pixels and was sampled uniformly at random from tissue-containing regions within all slides in a case. Early experiments with different patch sizes did not yield meaningful improvements. The CNN then extracted image-based features from the patches. A top-level average-pooling layer allowed the model to take as input different number of patches between training and evaluation. This enabled the use of a smaller number of patches and resultant higher case diversity during training, and a more extensive coverage of slides in each case with a larger number of patches during evaluation. A final logistic regression layer produced a prediction given the output of the average pooling layer.

*Survival Loss functions*

We experimented with three different loss functions for training the DLS. Early experiments (evaluated on the tune split) showed that the Censored Cross-Entropy described below gave the best results (Supplementary Figure S1) and was used for final model training.

The first tested loss function was based on the Cox partial likelihood[24], which is used for fitting Cox proportional hazard models but be extended to train neural networks as follows:

$$\max \prod_{i:O_i=1} \frac{e^{f(X_i)}}{\sum_{j:T_j \geq T_i} e^{f(X_j)}}$$

where $T_i$ is the event time or time of last follow-up, $O_i$ is an indicator variable for whether the event is observed, $X_i$ is the set of whole slide images and $f(X_i)$ is the DLS risk score, each for the i[th] case. In our implementation, we used Breslow's approximation[25] for handling tied event times. In principle, the loss for every single example is a function of all cases in the training data. In practice, we approximated the loss at each optimization step, by evaluating it over the examples in a small batch (n ≤ 128) instead of the entire training dataset.

Our second loss function was an exponential lower bound on the concordance index[26]. The concordance index is a common performance metric for survival models that corresponds to the probability that a randomly chosen pair of subjects is correctly ordered by the model in terms of event times. The concordance index itself is not differentiable, however, Vikas et al.[27] proposed the following differential lower bound that can be used for model optimization:

$$E := \{(i,j) | O_i = 1 \text{ and } T_j > T_i\}$$

$$\max \sum_{(i,j) \in E} 1 - e^{f(X_i) - f(X_j)}$$

Where $E$ is the set pairs of examples (i, j) where the i[th] event is observed and $T_j > T_i$. Similar to the Cox partial likelihood, we approximated this lower bound on the concordance index at each optimization step by evaluating it over the examples in a small batch (n ≤ 128) instead of the entire training dataset.

The final loss function, the censored cross-entropy, is an extension of the standard cross-entropy loss used for classification models to train survival prediction models with right-censored data. We modeled survival prediction as a classification problem instead of a regression or ranking problem, by discretizing time into intervals and training models to predict the discrete time interval in which the event occurred instead of a continuous event time or risk score. For examples with observed events, the standard cross-entropy was computed. However for censored examples, the time interval in which the event occurs is unknown. Therefore, we leverage the information that the event did not occur before the censorship time and maximize the log-likelihood of the event occurring in the interval of censorship or thereafter. The full loss function can be written as follows:

$$\max \sum_i \Big( O_i * \log(f(X_i)[Y_i]) + (1 - O_i) * \log(\sum_{y > Z_i} f(X_i)[y]) \Big)$$

Where $Y_i$ is the interval in which the event occurred (for example with observed events) and $Z_i$ is the latest interval whose endpoint is before the time of censorship (for censored examples). An important design consideration when using this loss function is how to discretize time. We used different percentiles of the time to death distribution for non-censored cases (i.e. quartiles). Discretization was done separately for each study to account for the considerable differences in survival times across studies (cancer types). To obtain a scalar risk score for evaluation, we took the negative of the expectation over the predicted time interval likelihood distribution. The negation ensured that higher risk score values indicates higher risk.

*Training Procedure*

Training examples consisted of sets of up to 16 image patches per case sampled uniformly from tissue across all the slides in that case. Tissue detection using a pixel-intensity-based threshold as well as data augmentation via stain normalization followed by color and orientation perturbations were both performed as described previously[9]. Training was performed using WSIs for both frozen and FFPE specimens. Numerical optimization of network parameters was done using the RMSProp optimizer[28] in TensorFlow in a distributed fashion, using 10 worker machines with 16 processors each.

For each study, we tuned the hyperparameters by randomly sampling 50 hyperparameter configurations and then training one model with each configuration for each of the 10 studies (500 models in total). The hyperparameter search space is detailed in Supplementary Table S2.

*Evaluation Procedure*

At evaluation we sampled 1024 patches per case, using the same procedure as during training. Early empirical studies using the tune set showed no performance benefit from sampling more patches. The final models used for evaluation were averaged in a number of ways. First, model weights were the exponential moving average of model weights across training steps, with a decay constant of 0.999. Next, instead of picking a single best training checkpoint (i.e. a model evaluated at particular training step) for each study, we used an ensemble of 50 checkpoints. Each model was trained for 500,000 steps and evaluated every 25,000 training steps, yielding 20 checkpoints per model, and a total of 1,000 checkpoints across 50 hyperparameter settings. The 50 checkpoints that achieved the highest c-index on the tune set were selected for the ensemble. The final ensemble prediction was the median of the 50 individual predictions.

**Survival analysis**

To avoid clinically irrelevant comparisons (e.g. 107 days vs 108 days), survival times were discretized from days to months for all analyses. For the Kaplan-Meier analysis, cases were first stratified into risk groups within each cancer type by choosing different risk score quantiles as thresholds. Stratification per cancer type is important because it ensures that the distribution of cancer types is the same across all risk groups. Without doing this it would have been possible to see differences in risk groups simply because one risk group contains more cases from cancers with a worse prognosis (e.g. OV vs BRCA). For the KM analysis by stage, we repeated the same procedure for cases from each stage.

We used Cox proportional hazards regression[29] as both an analytical tool and a predictive model. We used it first as an analytical tool for determining which variables were correlated with disease-specific survival by fitting multivariable models that include the DLS risk scores and baseline variables to our test dataset (Table 2). The pathologic stage was encoded as a numeric variable (i.e. 1, 2, 3, 4) in this analysis, because there were insufficient data for using a dummy variables for many studies. Age was also treated as a numeric variable. Age was divided by 10, so that the hazard ratio corresponds to the increased risk from an additional 10 years of age at the time of diagnosis. We fit a separate model for each study and a model across all studies combined. For the combined model, a dummy indicator variable for the cancer type was added.

In the second analysis, where we examined the additional prognostic value of adding the DLS to a multivariable model, we needed to control for the natural improvements in model fit with more input variables. Thus we used Cox regression as a predictive model, in conjunction with leave-one-out cross validation (LOO) across test set cases (Table 3). In this analysis, prognosis prediction performance was measured using the c-index, an extension of the AUC for binary outcomes without censorship[30]. Briefly, the concordance ("c") index is the number of "concordant" pairs (cases that were correctly ordered given the outcome and censorship time) divided by all informative pairs. Because different studies (cancer types) had markedly different followup periods and median survival times, the c-indices for the "combined" study summed concordant pairs and informative pairs solely within the same study. For example, the concordance index for the combined studies A and B was calculated as (concordant-pairs$_A$ + concordant-pairs$_B$) / (informative-pairs$_A$ + informative-pairs$_B$).

**Statistical analysis**

In Kaplan Meier analysis, survival curves for different groups were compared via a logrank test, implemented in the Lifelines python package (version 0.12.0)[31]. Confidence intervals and p-values for hazard ratios in Cox regression models were computed using Lifelines as well. Confidence intervals for the c-index and the delta in c-index between models were generated via bootstrapping with 9999 samples.

**Heatmap analysis**

Risk heatmaps for patch analysis were generated by running the DLS on a single patch at a time to produce patch-level DLS risk scores across entire slides. To generate visualizations for pathologist review, patches were sampled based on patch-level risk score from the top 25% and bottom 25% from each case. Patches were grouped by case and cases were organized by patient-level risk prediction. These organized patches were then reviewed by two pathologists to qualitatively evaluate high-level features that may be associated with both the case-level and patch-level risk scores.

## Results

**Comparing survival rates in low and high risk groups**

The output of the DLS is a continuous risk score that can be used as a feature for survival analysis. To define low and high risk groups, cases were binned into risk quartiles using DLS risk scores. Binning was done within each cancer type to ensure that the distribution of cancer types within each risk group was the same. A logrank test comparison between the Kaplan-Meier (KM) curves for the high and low risk groups yielded p<0.001 (Figure 2).

Given the known prognostic significance of stage, we assessed if the DLS could also sub-stratify patients' risk within each stage. The resulting Kaplan-Meier curves show that the DLS can further sub-stratify patients into low and high risk groups for stage II (p < 0.05) and stage III (p < 0.001), but not for stage I or stage IV (Figure 3).

**Multivariable analysis of the DLS and clinical metadata**

Next, we used multivariable Cox proportional-hazards regression to evaluate the significance of the DLS as a predictor of disease specific survival after adjusting for available variables: cancer stage, age, and sex. For the combined analysis including all 10 cancer types (i.e. "TCGA studies"), where cancer type was included as an indicator variable, the DLS was a significant predictor, with a hazard ratio of 1.48 (p<0.0001). To ensure that age and stage were adequately controlled for across studies, we further fit a combined model with additional interaction terms between the study and stage, and between study and age. In this expanded combined model, the p-value for the DLS remained below 0.001.

In subanalysis, for 5 of 10 cancer types, the DLS remained significantly associated with disease specific survival (Table 2; p=0.0002 to 0.0257). By contrast, cancer stage was a significant predictor in 7 studies, while age and sex were each significant predictors in only one study each.

**Measuring the added predictive value of the DLS**

The concordance index (or c-index) assesses the goodness-of-fit for survival model by calculating the probability of the model correctly ordering a (comparable) pair of cases in terms of their survival time. We compared the c-index of Cox-regression models with three different feature sets: (1) "DLS", consisting of the DLS predictions only; (2) "Baseline", consisting of stage, age, and sex; and (3) "Baseline+DLS", consisting of stage, age, sex, and DLS predictions. The c-index results for all cancer types combined and for each cancer type individually are summarized in Table 3. For the DLS model, the c-index for all 10 studies combined (comparisons across cases from different cancer types were excluded) was 61.1 (95% confidence interval (CI) [57.2, 65.1]). Within individual studies, the confidence intervals were too wide to draw meaningful conclusions due to low case volumes. We interpreted the delta in c-index between the "Baseline-only" and the "Baseline+DLS" models as a measure of the added predictive value of the DLS over the baseline variables. For all studies combined, the c-index delta was 3.7 (95% CI [1.0, 6.5]).

In addition to c-index, we also calculated the area under the receiver operating characteristic curve (AUC) for prediction of 5-year disease specific survival. Qualitatively similar results were observed, with the combined analysis showing an AUC improvement of 6.4 (95% CI [2.2, 10.8], Supplementary Table S4).

**Understanding the DLS**
To gain initial insights into the DLS, we first computed the correlation of the DLS predictions with the baseline variables of stage, TNM categories, and age. The DLS predictions were not correlated with age in any study, but were correlated with stage and T-category in several cancer types as well as in the combined analysis (Supplementary Table S5). Next, we analyzed the regions of each slide that contributed to the overall case classification by extracting the individual patches with the highest and lowest patch-level DLS risk scores for further review. Using KIRC as a representative example with a consistently high-performing DLS model, the patches with the "most confident" predictions for high or low risk tended primarily to contain tumor (Figure 4A-C), whereas patches with more intermediate prediction values tended to be non-tumor, such as fat, stroma, and fragmented tissue (Figure 4D). In this analysis, more detailed associations of histologic features and patch-level risk predictions were not identified.

## Discussion

Predicting patient prognosis in oncology underlies important clinical decisions regarding treatment and monitoring. In this work, we assessed the potential to improve predictions of disease-specific survival using a deep learning system trained without human annotations for known morphological features or regions of interest.

A natural question arises as to the value of developing algorithms to predict prognosis exclusively from learned features, versus leveraging region-level annotations such as tumor grade, nuclear pleomorphism, tumor-infiltrating lymphocytes, or mitotic figures among others. One straightforward advantage is to avoid the cost, tediousness, and difficulties associated with region-level annotations. Furthermore, the relatively unbiased nature of these weakly supervised models potentially enables the learning of previously unknown or unappreciated prognostic features. The primary disadvantage, on the other hand, was the increased number of cases required to train accurate models given that there was only a single case-level training label for each image, such as survival or disease progression. To place the difficulty of this problem in context, these labels correspond to $10^9$ pixels per image, often with several images per case, making for significantly weaker supervision than in typical image prediction tasks that deal with images sized $10^5$-$10^6$ pixels. In addition, cancer survival prediction is by nature limited to several orders of magnitude less data than typical image classification problems (e.g. $10^5$-$10^6$ images for ImageNet versus $10^2$-$10^3$ images here).

The DLS presented in this work learned morphologic features that were predictive of disease-specific survival in multiple cancer types. While we did not identify any clear trends or confounders specific to the cancer types for which the models performed best, future work to better understand the effects of sample size, image-specific variables, and disease-specific variables on clinical predictions from WSIs will be important for the field. Our solution for weak supervision involves a neural network architecture that randomly samples multiple tissue-containing patches for each case at training time. This sampling approach has three main advantages. First, it provides a high probability of seeing patches containing informative features in each training iteration, and even more so across training iterations. Second, assuming each case contains more than one informative image patch, it substantially expands the effective dataset size by increasing the diversity of examples. Third, even uninformative patches have a regularization effect on the training. A similar approach has been explored[18] though only for tissue microarrays of a single cancer type and using image features from a frozen model that was trained on ImageNet. We have provided a more comprehensive analysis than prior work by developing and validating our DLS models across multiple cancer types on WSIs without region of interest annotations.

In our study, the fact that the DLS output remained significantly associated with disease specific survival even after adjusting for age and cancer stage suggests that the DLS learned prognostic morphologic features that were independent from these baseline variables. In an effort to better understand some of the learned features, we applied the DLS to every image patch on each slide to obtain "patch-level prognosis estimates" across the entire image. In this analysis, the most confident prognostic regions were comprised primarily of tumor with minimal intervening stroma or other obvious histological structures. While other machine learning efforts have identified prognostic significance for non-tumor elements[17,32], our observations suggest that at least for our specific models, the morphologic features of the tumor appear to be more relevant than non-tumor regions. However, elucidating the morphological features that the DLS learned to help distinguish between high risk and low risk cases remains an exciting but challenging topic for future efforts, and one that will likely require identification of unique features for different tumor types. One intriguing hypothesis is that DLS-learned features may correspond to previously unappreciated representations of tumor biology in the histology, and that underlying biological pathways or molecular mechanisms may be further elucidated via focused evaluation of regions highlighted by the DLS.

Though we have presented promising results for a challenging deep learning problem, there are several notable limitations to our study. First, despite leveraging data across 10 cancer types from the biggest public dataset available (TCGA), each cancer type's test dataset contained fewer than 250 cases and fewer than 100 disease specific survival events, resulting in wide confidence intervals that limit statistical conclusions (and highlight the importance of reporting model performance confidence intervals when publishing in this field). As such, this work represents a proof-of-concept study to refine methods and to better understand the feasibility of weakly supervised, direct clinical outcome prediction. While the models did learn prognostic signals, these findings require additional development and validation in larger datasets to further improve predictions and more accurately estimate effect sizes, let alone to demonstrate clinical value. Second, our methods and results are limited to datasets from TCGA, for which there are typically a small number of images per case and tumor purity in each image is high[14]. Thus it remains to be seen if the random "patch-sampling" approach described here will be effective in real-world clinical settings where tumor purity is more variable, sectioning protocols may differ, and many slides are typically available for each case. Additionally, while the possible confounding effect of treatment differences between patients were not addressed in these data, all of the patients in these studies were untreated at the time of tissue sampling and the risk stratification on baseline variables shows the expected pattern despite possible differences in treatment. We also note that the DLS was only presented with regions of primary, untreated tumors (as per TCGA inclusion criteria and sampling). While this potentially allowed learning of features associated with the primary tumor such as tumor invasion or grade, the DLS is arguably less likely to have learned features associated with additional specimens such as lymph nodes, margin regions, or metastatic sites. Indeed, the DLS predictions did correlate with the "T" categorization of the TNM staging in the combined analysis, but not with the "N" categorization (Supplementary Table S5). Future work using additional slides may be able to further inform risk stratification via learning of additional histological features. Lastly, this work does not specifically incorporate available molecular information from TCGA, which would likely require cancer type-specific molecular analyses and larger datasets.

In conclusion, we have demonstrated promising results for direct prediction of clinical outcomes in a weakly-supervised setting, without the use of any region-level expert-annotations for training. We hope this work provides useful insights and benchmarks regarding dataset requirements and modeling approaches for survival prediction, especially as it relates to use of the publicly available TCGA data.

**Acknowledgements**
We thank our colleagues at Google Health for their support of this work. We would also like to thank Akinori Mitani for critical review of the manuscript, and Angela Lin and Yuannan Cai for help in acquiring and organizing the publicly available data.

# References

1.  Weiser MR, Gönen M, Chou JF, Kattan MW, Schrag D. Predicting survival after curative colectomy for cancer: individualizing colon cancer staging. J Clin Oncol. 2011;29: 4796–4802.

2.  Cooperberg MR, Pasta DJ, Elkin EP, Litwin MS, Latini DM, Du CHANE J, et al. The University of California, San Francisco Cancer of the Prostate Risk Assessment score: a straightforward and reliable preoperative predictor of disease recurrence after radical prostatectomy. Journal of Urology. 2005. pp. 1938–1942. doi:10.1097/01.ju.0000158155.33890.e7

3.  Sparano JA, Gray RJ, Ravdin PM, Makower DF, Pritchard KI, Albain KS, et al. Clinical and Genomic Risk to Guide the Use of Adjuvant Therapy for Breast Cancer. N Engl J Med. 2019. doi:10.1056/NEJMoa1904819

4.  Sparano JA, Gray RJ, Makower DF, Pritchard KI, Albain KS, Hayes DF, et al. Adjuvant Chemotherapy Guided by a 21-Gene Expression Assay in Breast Cancer. N Engl J Med. 2018;379: 111–121.

5.  Rakha EA, El-Sayed ME, Lee AHS, Elston CW, Grainge MJ, Hodi Z, et al. Prognostic significance of Nottingham histologic grade in invasive breast carcinoma. J Clin Oncol. 2008;26: 3153–3158.

6.  LeCun Y, Bengio Y, Hinton G. Deep learning. Nature. 2015;521: 436–444.

7.  Gulshan V, Peng L, Coram M, Stumpe MC, Wu D, Narayanaswamy A, et al. Development and Validation of a Deep Learning Algorithm for Detection of Diabetic Retinopathy in Retinal Fundus Photographs. JAMA. 2016;316: 2402–2410.

8.  Ehteshami Bejnordi B, Veta M, Johannes van Diest P, van Ginneken B, Karssemeijer N, Litjens G, et al. Diagnostic Assessment of Deep Learning Algorithms for Detection of Lymph Node Metastases in Women With Breast Cancer. JAMA. 2017;318: 2199–2210.

9.  Liu Y, Kohlberger T, Norouzi M, Dahl GE, Smith JL, Mohtashamian A, et al. Artificial Intelligence-Based Breast Cancer Nodal Metastasis Detection. Arch Pathol Lab Med. 2018. doi:10.5858/arpa.2018-0147-OA

10. Nagpal K, Foote D, Liu Y, Chen P-HC, Wulczyn E, Tan F, et al. Development and validation of a deep learning algorithm for improving Gleason scoring of prostate cancer. npj Digital Medicine. 2019;2: 48.

11. Saltz J, Gupta R, Hou L, Kurc T, Singh P, Nguyen V, et al. Spatial Organization and Molecular Correlation of Tumor-Infiltrating Lymphocytes Using Deep Learning on Pathology Images. Cell Rep. 2018;23: 181–193.e7.

12. Mobadersany P, Yousefi S, Amgad M, Gutman DA, Barnholtz-Sloan JS, Velázquez Vega JE, et al. Predicting cancer outcomes from histology and genomics using convolutional networks. Proc Natl Acad Sci U S A. 2018;115: E2970–E2979.

13. Yu K-H, Zhang C, Berry GJ, Altman RB, Ré C, Rubin DL, et al. Predicting non-small cell lung cancer prognosis by fully automated microscopic pathology image features. Nat

Commun. 2016;7: 12474.

14. Cooper LAD, Demicco EG, Saltz JH, Powell RT, Rao A, Lazar AJ. PanCancer insights from The Cancer Genome Atlas: the pathologist's perspective. The Journal of Pathology. 2018. pp. 512–524. doi:10.1002/path.5028

15. Wang C, Pécot T, Zynger DL, Machiraju R, Shapiro CL, Huang K. Identifying survival associated morphological features of triple negative breast cancer using multiple datasets. J Am Med Inform Assoc. 2013;20: 680–687.

16. Yu K-H, Berry GJ, Rubin DL, Ré C, Altman RB, Snyder M. Association of Omics Features with Histopathology Patterns in Lung Adenocarcinoma. Cell Syst. 2017;5: 620–627.e3.

17. Kather JN, Krisam J, Charoentong P, Luedde T, Herpel E, Weis C-A, et al. Predicting survival from colorectal cancer histology slides using deep learning: A retrospective multicenter study. PLoS Med. 2019;16: e1002730.

18. Bychkov D, Linder N, Turkki R, Nordling S, Kovanen PE, Verrill C, et al. Deep learning based tissue analysis predicts outcome in colorectal cancer. Sci Rep. 2018;8: 3395.

19. Zhu X, Yao J, Zhu F, Huang J. WSISA: Making Survival Prediction from Whole Slide Histopathological Images. 2017 IEEE Conference on Computer Vision and Pattern Recognition (CVPR). 2017. doi:10.1109/cvpr.2017.725

20. Weinstein JN, The Cancer Genome Atlas Research Network, Collisson EA, Mills GB, Mills Shaw KR, Ozenberger BA, et al. The Cancer Genome Atlas Pan-Cancer analysis project. Nat Genet. 2013;45: 1113–1120.

21. Liu J, Lichtenberg T, Hoadley KA, Poisson LM, Lazar AJ, Cherniack AD, et al. An Integrated TCGA Pan-Cancer Clinical Data Resource to Drive High-Quality Survival Outcome Analytics. Cell. 2018;173: 400–416.e11.

22. The Cancer Genome Atlas Network. Genomic Classification of Cutaneous Melanoma. Cell. 2015;161: 1681.

23. Howard AG, Zhu M, Chen B, Kalenichenko D, Wang W, Weyand T, et al. MobileNets: Efficient Convolutional Neural Networks for Mobile Vision Applications. arXiv [cs.CV]. 2017. Available: http://arxiv.org/abs/1704.04861

24. Cox DR. Partial Likelihood. Biometrika. 1975. p. 269. doi:10.2307/2335362

25. Breslow N. Covariance Analysis of Censored Survival Data. Biometrics. 1974. p. 89. doi:10.2307/2529620

26. Harrell FE Jr, Lee KL, Mark DB. Multivariable prognostic models: issues in developing models, evaluating assumptions and adequacy, and measuring and reducing errors. Stat Med. 1996;15: 361–387.

27. Steck H, Krishnapuram B, Dehing-oberije C, Lambin P, Raykar VC. On Ranking in Survival Analysis: Bounds on the Concordance Index. In: Platt JC, Koller D, Singer Y, Roweis ST, editors. Advances in Neural Information Processing Systems 20. Curran Associates, Inc.; 2008. pp. 1209–1216.

28. Tieleman T, Hinton G. Lecture 6.5-rmsprop: Divide the gradient by a running average of its recent magnitude. COURSERA: Neural networks for machine learning. 2012;4: 26–31.

29. Cox DR. Regression Models and Life-Tables. J R Stat Soc Series B Stat Methodol. 1972;34: 187–220.

30. Harrell FE. Evaluating the yield of medical tests. JAMA: The Journal of the American Medical Association. 1982. pp. 2543–2546. doi:10.1001/jama.247.18.2543

31. Davidson-Pilon C, Kalderstam J, Zivich P, Kuhn B, Fiore-Gartland A, Moneda L, et al. CamDavidsonPilon/lifelines: v0.21.3. 2019. doi:10.5281/zenodo.3240536

32. Beck AH, Sangoi AR, Leung S, Marinelli RJ, Nielsen TO, van de Vijver MJ, et al. Systematic analysis of breast cancer morphology uncovers stromal features associated with survival. Sci Transl Med. 2011;3: 108ra113.

# Tables

**Table 1. Dataset Overview.** Our datasets were derived from The Cancer Genome Atlas (TCGA). Cases with known disease specific survival (DSS), pathologic stage, age, and sex were assigned into train, tune, and test splits in a ratio of 2:1:1. Each TCGA study code refers to a cancer type, and "Combined" refers to all 10 studies combined. Cancer stage distribution is presented in Supplementary Table S1.

| Study | Cases | | | DSS Events (%) | | | Slides | | |
|---|---|---|---|---|---|---|---|---|---|
| | Train | Tune | Test | Train | Tune | Test | Train | Tune | Test |
| BLCA (bladder urothelial carcinoma) | 197 | 98 | 96 | 62 (31%) | 31 (32%) | 30 (31%) | 437 | 205 | 206 |
| BRCA (breast invasive carcinoma) | 488 | 247 | 250 | 40 (8%) | 19 (8%) | 20 (8%) | 1182 | 599 | 631 |
| COAD (colon adenocarcinoma) | 218 | 110 | 103 | 32 (15%) | 16 (15%) | 13 (13%) | 625 | 313 | 310 |
| HNSC (head and neck squamous cell carcinoma) | 196 | 99 | 101 | 52 (27%) | 27 (27%) | 28 (28%) | 481 | 250 | 247 |
| KIRC (kidney renal clear cell carcinoma) | 260 | 130 | 130 | 55 (21%) | 27 (21%) | 27 (21%) | 777 | 395 | 382 |
| LIHC (liver hepatocellular carcinoma) | 165 | 83 | 85 | 32 (19%) | 17 (20%) | 18 (21%) | 341 | 165 | 172 |
| LUAD (lung adenocarcinoma) | 233 | 115 | 112 | 54 (23%) | 28 (24%) | 26 (23%) | 619 | 283 | 282 |
| LUSC (lung squamous cell carcinoma) | 219 | 108 | 109 | 45 (21%) | 22 (20%) | 21 (19%) | 542 | 275 | 269 |
| OV (ovarian serous cystadenocarcinoma) | 272 | 133 | 137 | 151 (56%) | 73 (55%) | 76 (55%) | 607 | 299 | 298 |
| STAD (stomach adenocarcinoma) | 198 | 95 | 93 | 48 (24%) | 24 (25%) | 25 (27%) | 464 | 227 | 212 |
| Combined | 2446 | 1218 | 1216 | 571 (23%) | 284 (23%) | 284 (23%) | 6075 | 3011 | 3009 |

**Table 2. Multivariable Cox proportional hazards regression analysis demonstrates association of the deep learning system (DLS) with disease-specific survival.** Each column header represents one of the input variables for the multivariable analysis, with HR indicating the hazard ratio. For the combined analysis, the study was also included as an indicator variable (coefficients not shown). Univariable analysis is presented in Supplementary Table S2.

| Study | Risk Factor | | | | | | | |
|---|---|---|---|---|---|---|---|---|
| | DLS | | Age | | Male | | Stage | |
| | HR | p | HR | p | HR | p | HR | p |
| BLCA | 0.75 [0.45, 1.24] | 0.2636 | 1.27 [0.82, 1.98] | 0.2809 | 1.53 [0.57, 4.11] | 0.3939 | **2.30 [1.37, 3.86]** | **0.0016** |
| BRCA | **2.86 [1.42, 5.76]** | **0.0034** | 1.01 [0.73, 1.40] | 0.9412 | NaN | NaN | 1.72 [0.94, 3.12] | 0.0767 |
| COAD | **4.03 [1.92, 8.44]** | **0.0002** | 0.85 [0.53, 1.38] | 0.5086 | 1.18 [0.38, 3.69] | 0.7769 | **11.86 [4.18, 33.66]** | **0.0000** |
| HNSC | **2.32 [1.11, 4.88]** | **0.0257** | 0.93 [0.63, 1.39] | 0.7338 | 0.91 [0.37, 2.20] | 0.8262 | **2.26 [1.16, 4.42]** | **0.0171** |
| KIRC | **1.88 [1.23, 2.87]** | **0.0035** | 0.99 [0.69, 1.42] | 0.9517 | **0.33 [0.14, 0.77]** | **0.0107** | **3.20 [2.02, 5.07]** | **0.0000** |
| LIHC | **2.74 [1.54, 4.86]** | **0.0006** | 1.23 [0.84, 1.82] | 0.2869 | 0.99 [0.32, 3.03] | 0.9809 | **2.31 [1.25, 4.24]** | **0.0072** |
| LUAD | 1.35 [0.87, 2.08] | 0.1824 | 0.78 [0.56, 1.10] | 0.1546 | 1.36 [0.58, 3.17] | 0.4762 | **2.11 [1.50, 2.97]** | **0.0000** |
| LUSC | 1.97 [0.90, 4.32] | 0.0894 | 0.83 [0.49, 1.39] | 0.4785 | 1.49 [0.54, 4.14] | 0.4404 | 1.48 [0.91, 2.41] | 0.1162 |
| OV | 1.24 [0.95, 1.63] | 0.1157 | **1.26 [1.02, 1.55]** | **0.0326** | NaN | NaN | 1.45 [0.95, 2.20] | 0.0845 |
| STAD | 1.50 [0.85, 2.62] | 0.1602 | 0.96 [0.69, 1.35] | 0.8318 | 1.94 [0.79, 4.76] | 0.1496 | **2.19 [1.26, 3.83]** | **0.0058** |
| Combined | **1.48 [1.28, 1.70]** | **<0.0001** | 1.07 [0.96, 1.18] | 0.2221 | 1.08 [0.80, 1.48] | 0.6063 | **2.30 [1.99, 2.66]** | **<0.0001** |

**Table 3. C-index for Cox regression models using DLS and baseline variables as input.**
(1) deep learning system ("DLS-only"), (2) stage, age, sex ("Baseline-only"), or (3) age, stage, sex, and DLS ("Baseline + DLS"). Significant differences based on confidence intervals are highlighted in bold.

| Study | DLS (1) | Baseline (2) | Baseline + DLS (3) | Delta (3 -2) |
|---|---|---|---|---|
| BLCA | 54.0 [43.3, 64.8] | 69.0 [57.4, 80.8] | 68.3 [56.0, 80.1] | -0.7 [-4.6, 2.8] |
| BRCA | 72.0 [55.5, 87.3] | 64.3 [45.6, 78.3] | 71.0 [53.8, 85.7] | 6.7 [-7.9, 20.6] |
| COAD | 70.9 [54.0, 85.4] | 80.0 [66.6, 90.9] | 91.9 [85.7, 96.6] | **11.9 [3.8, 23.1]** |
| HNSC | 58.2 [46.0, 70.0] | 49.7 [38.8, 60.4] | 64.3 [52.6, 75.0] | **14.5 [5.3, 24.6]** |
| KIRC | 71.1 [59.4, 82.5] | 85.7 [77.9, 92.4] | 85.9 [79.1, 92.2] | 0.2 [-3.1, 4.0] |
| LIHC | 71.3 [53.8, 88.0] | 77.3 [62.3, 88.5] | 80.1 [67.6, 91.2] | 2.8 [-6.1, 12.6] |
| LUAD | 46.3 [32.5, 59.8] | 75.4 [65.1, 84.1] | 74.8 [64.6, 83.8] | -0.6 [-4.0, 2.7] |
| LUSC | 62.1 [47.3, 75.4] | 55.7 [42.6, 68.6] | 60.3 [46.4, 72.7] | 4.6 [-8.7, 15.8] |
| OV | 53.9 [45.2, 62.5] | 60.3 [52.6, 67.7] | 61.3 [53.4, 68.7] | 1.0 [-3.4, 5.8] |
| STAD | 68.7 [57.8, 78.3] | 67.5 [57.5, 77.4] | 72.4 [62.3, 81.9] | 4.9 [-2.1, 12.0] |
| Combined | 61.1 [57.2, 65.1] | 66.9 [63.1, 70.8] | 70.6 [67.1, 74.2] | **3.7 [1.0, 6.5]** |

# Figures

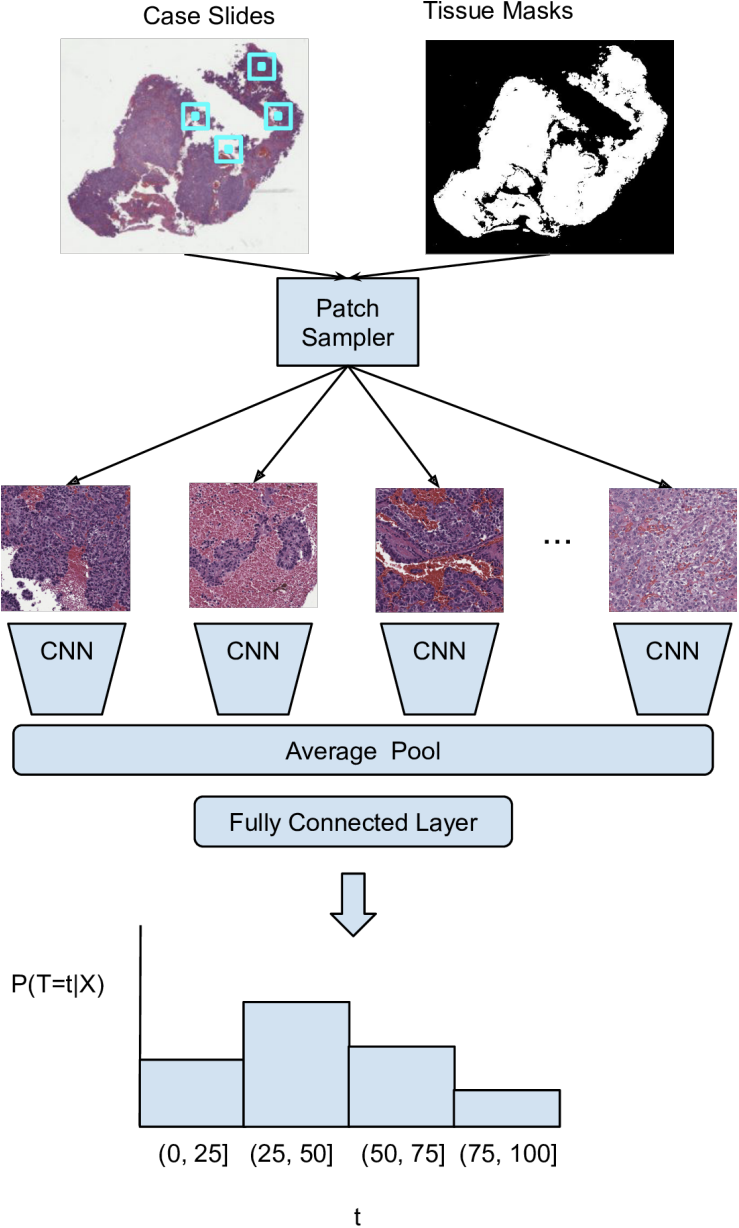

**Figure 1. Summary of the weakly supervised learning approach for directly predicting disease specific survival.** For each case, cropped image patches were uniformly sampled from tissue-containing areas across all slides available for a given case. Next, image features were extracted for each patch by a convolutional neural network (CNN). These patch-level features were averaged (on a per-channel basis) and fed to a fully connected layer. Our custom loss function divided the follow-up period into four discrete bins depending on right-censorship time and outcome (Methods). As such, the model was designed to output a probability distribution over discretized survival times.

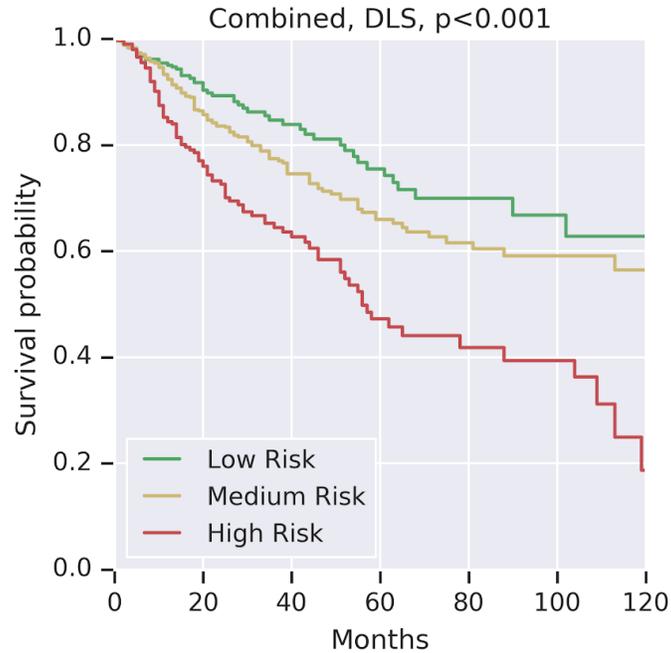

**Figure 2. Kaplan Meier curves for DLS risk groups.** To define low and high risk groups, cases were binned into risk quartiles using DLS risk scores. Binning was done within each cancer type to ensure that the distribution of cancer types within each risk group was the same. Different colors represent the different risk groups: green for the low risk (0th -25th percentile); yellow for medium risk (25th-75th percentile), and red for high risk (75th-100th percentile). P-values were calculated using the binary logrank test comparing the low and high risk groups.

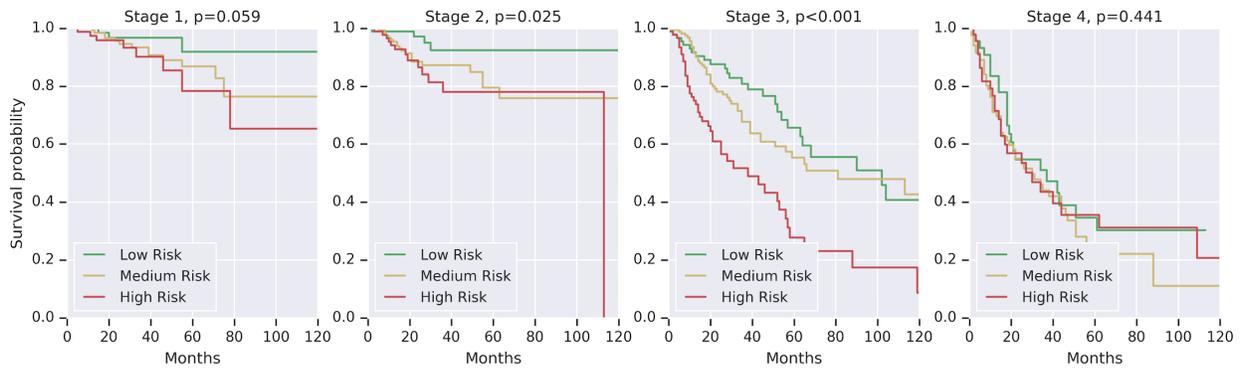

**Figure 3. Kaplan Meier curves for DLS risk groups within each cancer stage.** To define low and high risk groups, cases were binned into risk quartiles using DLS risk scores. Binning was done within each stage and cancer type. This ensures that for each stage, the distribution of cancer types within each risk group was the same. P-values were calculated using the binary logrank test comparing the low and high risk groups.

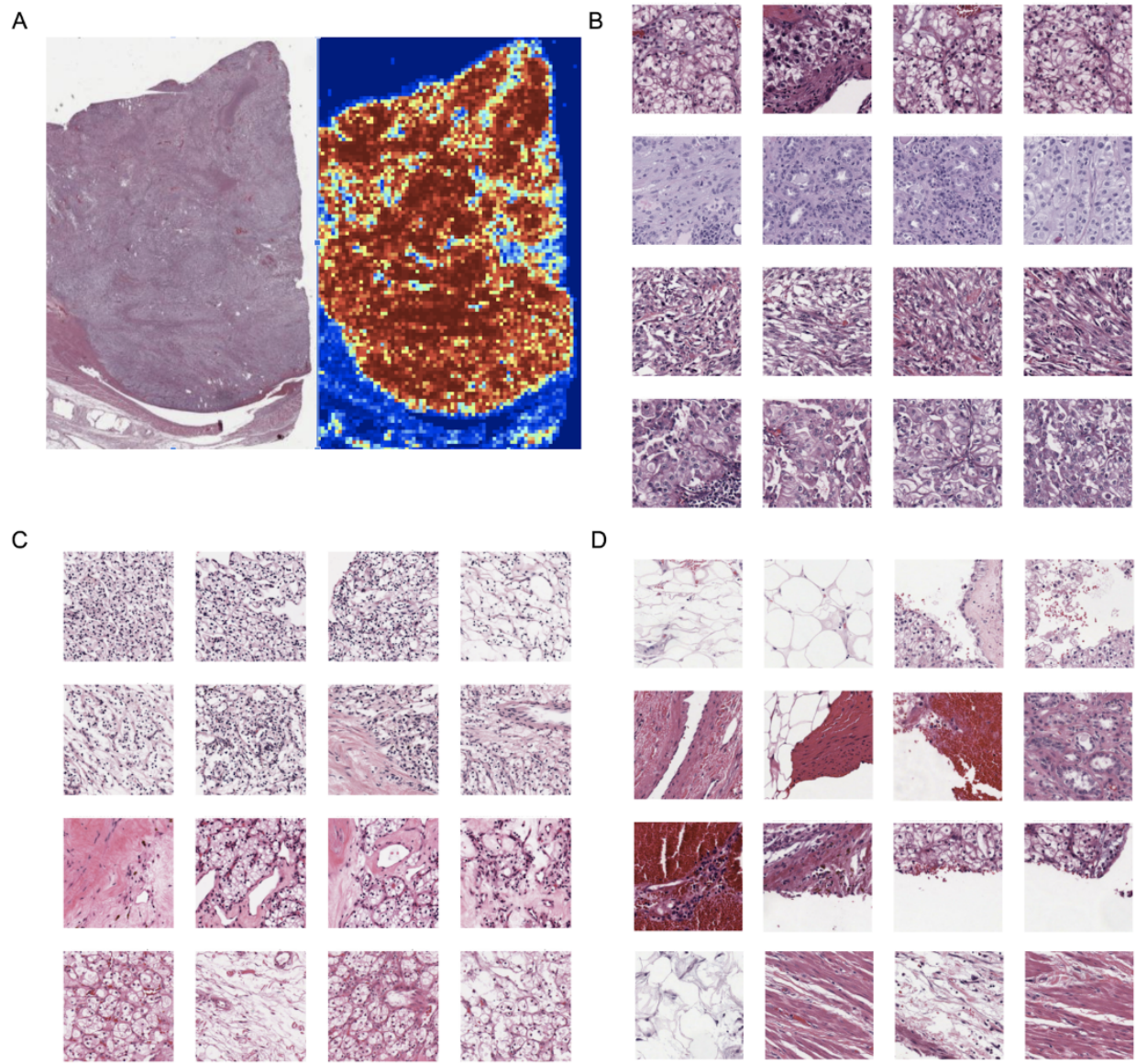

**Figure 4. Visualization of image patches influencing survival prediction.** (A) Example of WSI kidney renal clear cell carcinoma (KIRC) predicted to be high risk (left), with the DLS-predicted "risk heatmap" on the right; red patches correspond to "high-risk" and blue patches to "low-risk" patch-level predictions (Methods). (B) "Highest-risk" patches from cases predicted to be high-risk. (C) "Lowest-risk" patches from cases predicted to be low-risk. (D) "Lowest-risk" patches from cases predicted to be high-risk. For B, C, and D, patches in the same row are from the same case and each row represents a different case.

# Supplementary Information

**Supplementary Table S1. Pathologic Stage distribution for each study.** *only clinical stage (rather than pathologic stage) was available and used for OV.

| Study | Stage | | | | | | | | | | | |
|---|---|---|---|---|---|---|---|---|---|---|---|---|
| | Train | | | | Tune | | | | Test | | | |
| | I | II | III | IV | I | II | III | IV | I | II | III | IV |
| BLCA | 1 (0.5%) | 73 (37.1%) | 62 (31.5%) | 61 (31.0%) | 1 (1.0%) | 26 (26.5%) | 35 (35.7%) | 36 (36.7%) | 0 (0.0%) | 29 (30.2%) | 35 (36.5%) | 32 (33.3%) |
| BRCA | 76 (15.6%) | 286 (58.6%) | 118 (24.2%) | 8 (1.6%) | 45 (18.2%) | 145 (58.7%) | 52 (21.1%) | 5 (2.0%) | 53 (21.2%) | 139 (55.6%) | 52 (20.8%) | 6 (2.4%) |
| COAD | 33 (15.1%) | 84 (38.5%) | 65 (29.8%) | 36 (16.5%) | 24 (21.8%) | 40 (36.4%) | 33 (30.0%) | 13 (11.8%) | 19 (18.4%) | 44 (42.7%) | 25 (24.3%) | 15 (14.6%) |
| HNSC | 11 (5.6%) | 33 (16.8%) | 30 (15.3%) | 122 (62.2%) | 5 (5.1%) | 15 (15.2%) | 21 (21.2%) | 58 (58.6%) | 6 (5.9%) | 15 (14.9%) | 21 (20.8%) | 59 (58.4%) |
| KIRC | 132 (50.8%) | 29 (11.2%) | 61 (23.5%) | 38 (14.6%) | 67 (51.5%) | 15 (11.5%) | 27 (20.8%) | 21 (16.2%) | 66 (50.8%) | 12 (9.2%) | 32 (24.6%) | 20 (15.4%) |
| LIHC | 79 (47.9%) | 46 (27.9%) | 39 (23.6%) | 1 (0.6%) | 43 (51.8%) | 21 (25.3%) | 19 (22.9%) | 0 (0.0%) | 47 (55.3%) | 13 (15.3%) | 23 (27.1%) | 2 (2.4%) |
| LUAD | 117 (50.2%) | 57 (24.5%) | 47 (20.2%) | 12 (5.2%) | 69 (60.0%) | 29 (25.2%) | 12 (10.4%) | 5 (4.3%) | 69 (61.6%) | 25 (22.3%) | 13 (11.6%) | 5 (4.5%) |
| LUSC | 113 (51.6%) | 63 (28.8%) | 40 (18.3%) | 3 (1.4%) | 54 (50.0%) | 36 (33.3%) | 17 (15.7%) | 1 (0.9%) | 51 (46.8%) | 42 (38.5%) | 13 (11.9%) | 3 (2.8%) |
| OV* | 4 (1.5%) | 19 (7.0%) | 204 (75.0%) | 45 (16.5%) | 3 (2.3%) | 6 (4.5%) | 108 (81.2%) | 16 (12.0%) | 8 (5.8%) | 4 (2.9%) | 102 (74.5%) | 23 (16.8%) |
| STAD | 28 (14.1%) | 68 (34.3%) | 83 (41.9%) | 19 (9.6%) | 13 (13.7%) | 26 (27.4%) | 48 (50.5%) | 8 (8.4%) | 12 (12.9%) | 30 (32.3%) | 39 (41.9%) | 12 (12.9%) |
| Combined | 594 (24.3%) | 758 (31.0%) | 749 (30.6%) | 345 (14.1%) | 324 (26.6%) | 359 (29.5%) | 372 (30.5%) | 163 (13.4%) | 331 (27.2%) | 353 (29.0%) | 355 (29.2%) | 177 (14.6%) |

**Supplementary Table S2. Hyperparameter search space (see Methods for usage).**

| Hyperparameter | Description | Values |
| --- | --- | --- |
| Fixation Types | Fixation types for slides for both training and evaluation | FFPE, "FFPE and FROZEN" |
| Patch size | Height and width of each image patch | 256 |
| Patch set size | Number of patches sampled from a case to form a single training example: | 1, 4, 8, 16 |
| Magnification | Image magnification at which the patches are extracted | 20X, 10X, 5X |
| Number of layers | Number of layers used in our MobileNet-based architecture | 4, 8, 12 |
| Base depth | Depth of the first convolution layer in the MobilNet CNN; depth grows by a factor of 1.25 for every 2 layers in the network. | 8, 16 |
| L2 regularization weight | Weight of the L2 loss used for regularization | 0.004, 0.0004, 0.00004, 0.000004 |
| Initial Learning rate | Initial learning rate used for the RMSPROP optimizer; decay rate was 0.99 every 20,000 steps. | 0.005, 0.001, 0.0005, 0.0001 |
| Thresholds | Percentile thresholds used for binning time in the censored cross-entropy loss. | [50], [25, 75], [25, 50, 75] |
| Training dataset | Trained on study (cancer-type) specific data, or combined across all cancers | Combined or cancer-specific |

**Supplementary Table S3. Univariable Cox analysis (see Table 2 for multivariable analysis).**

| Study | Risk Factor | | | | | | | |
|---|---|---|---|---|---|---|---|---|
| | DLS | | Age | | Male | | Stage | |
| | HR | p | HR | p | HR | p | HR | p |
| BLCA | 0.81 [0.50, 1.32] | 0.3935 | 1.31 [0.91, 1.90] | 0.1516 | 1.41 [0.54, 3.70] | 0.4801 | **2.48 [1.47, 4.20]** | **0.0007** |
| BRCA | **3.86 [2.15, 6.96]** | **0.0000** | 0.99 [0.71, 1.38] | 0.9509 | NaN | NaN | **2.62 [1.54, 4.47]** | **0.0004** |
| COAD | **2.09 [1.21, 3.63]** | **0.0087** | 0.68 [0.45, 1.03] | 0.0706 | 0.99 [0.33, 2.96] | 0.9827 | **6.66 [2.76, 16.09]** | **0.0000** |
| HNSC | 1.82 [0.98, 3.40] | 0.0583 | 0.96 [0.66, 1.41] | 0.8406 | 1.07 [0.47, 2.45] | 0.8706 | **1.98 [1.08, 3.64]** | **0.0274** |
| KIRC | **2.82 [1.85, 4.32]** | **0.0000** | 1.14 [0.86, 1.52] | 0.3466 | 0.57 [0.27, 1.22] | 0.1500 | **3.22 [2.12, 4.90]** | **0.0000** |
| LIHC | **3.26 [1.93, 5.53]** | **0.0000** | 0.97 [0.70, 1.35] | 0.8658 | 0.90 [0.35, 2.32] | 0.8298 | **2.60 [1.52, 4.42]** | **0.0004** |
| LUAD | 1.06 [0.68, 1.65] | 0.7954 | 0.78 [0.56, 1.09] | 0.1464 | 1.35 [0.62, 2.97] | 0.4505 | **2.06 [1.48, 2.85]** | **0.0000** |
| LUSC | 1.98 [0.99, 3.94] | 0.0517 | 0.93 [0.58, 1.48] | 0.7615 | 1.69 [0.62, 4.62] | 0.3075 | **1.68 [1.05, 2.68]** | **0.0291** |
| OV | 1.19 [0.90, 1.55] | 0.2168 | 1.22 [0.98, 1.50] | 0.0703 | NaN | NaN | 1.35 [0.90, 2.03] | 0.1446 |
| STAD | 1.74 [1.05, 2.90] | 0.0329 | 0.89 [0.64, 1.25] | 0.5090 | 1.86 [0.78, 4.47] | 0.1636 | **2.26 [1.34, 3.82]** | **0.0023** |

**Supplementary Table S4. AUC for binarized 5-year disease-specific survival (instead of c-index as in Table 3).** In STAD, only 3 cases had at least 5 years of follow-up and survived for at least 5 years.

| Study | DLS (1) | Baseline (2) | Baseline + DLS (3) | Delta (3 -2) |
|---|---|---|---|---|
| BLCA | 53.1 [31.9, 73.8] | 68.3 [46.2, 87.7] | 69.9 [50.3, 87.3] | 1.6 [-9.5, 14.2] |
| BRCA | 74.0 [58.7, 87.6] | 58.4 [37.3, 75.7] | 70.0 [51.3, 85.6] | **11.6 [-1.5, 25.0]** |
| COAD | 87.5 [66.7, 100.0] | 72.9 [43.3, 97.3] | 94.8 [80.0, 100.0] | **21.9 [2.7, 41.7]** |
| HNSC | 53.5 [33.5, 73.8] | 39.9 [15.4, 66.7] | 60.1 [38.2, 82.5] | 20.2 [-1.3, 44.8] |
| KIRC | 72.8 [59.4, 85.1] | 84.5 [73.4, 94.1] | 87.7 [77.1, 96.3] | 3.2 [-1.3, 8.4] |
| LIHC | 77.6 [56.1, 93.8] | 65.6 [41.7, 87.4] | 74.1 [52.4, 91.8] | 8.5 [-9.4, 27.0] |
| LUAD | 50.9 [29.6, 70.8] | 66.3 [39.4, 89.6] | 67.1 [40.8, 90.6] | 0.9 [-2.9, 6.2] |
| LUSC | 61.1 [36.7, 82.0] | 65.6 [45.5, 84.0] | 76.1 [56.0, 92.9] | 10.5 [-8.1, 30.5] |
| OV | 59.9 [47.6, 72.6] | 56.6 [43.1, 68.6] | 60.5 [46.3, 72.2] | 4.0 [-3.3, 11.3] |
| STAD | 28.0 [0.0, 65.5] | 64.0 [4.1, 100.0] | 51.3 [4.1, 100.0] | -12.7 [-42.4, 0.0] |
| Combined | 64.3 [58.0, 70.3] | 63.7 [57.1, 70.8] | 70.1 [63.8, 76.8] | **6.4 [2.2, 10.8]** |

**Supplementary Table S5. Correlation of the DLS predictions with clinical variables.**
Spearman's rank correlation and p-values (with significant values in bold) for correlation of each variable with DLS predictions across cancer types. The Spearman's rank correlation coefficient was used to account for different numerical scales of each variable.

|          | Stage          | T              | N              | M              | Age            |
|----------|----------------|----------------|----------------|----------------|----------------|
| BLCA     | 7.54 (0.465)   | **15.82 (0.124)** | 6.69 (0.517)   | 15.50 (0.132)  | 6.79 (0.511)   |
| BRCA     | **24.39 (0.000)** | **31.00 (0.000)** | 12.23 (0.053)  | 12.16 (0.055)  | -7.39 (0.244)  |
| COAD     | 0.39 (0.969)   | -0.73 (0.942)  | 3.80 (0.703)   | 2.90 (0.771)   | 3.16 (0.751)   |
| HNSC     | 6.42 (0.524)   | 14.92 (0.136)  | -1.90 (0.851)  | **-1.54 (0.878)** | -0.94 (0.926)  |
| KIRC     | **29.07 (0.001)** | **27.77 (0.001)** | -13.34 (0.130) | 5.39 (0.542)   | 14.11 (0.109)  |
| LIHC     | **23.01 (0.034)** | **24.85 (0.022)** | -11.14 (0.310) | -5.20 (0.636)  | -12.73 (0.246) |
| LUAD     | -3.30 (0.730)  | 7.41 (0.438)   | -4.82 (0.614)  | 17.33 (0.068)  | 8.50 (0.373)   |
| LUSC     | 8.91 (0.357)   | 17.99 (0.061)  | 3.75 (0.699)   | -5.09 (0.599)  | **20.55 (0.032)** |
| OV       | -5.81 (0.500)  | N/A            | N/A            | N/A            | N/A            |
| STAD     | **22.21 (0.032)** | -2.08 (0.843)  | 8.26 (0.431)   | **27.00 (0.009)** | -5.93 (0.572)  |
| Combined | **15.55 (0.000)** | **18.95 (0.000)** | 1.59 (0.602)   | 5.46 (0.073)   | 5.34 (0.079)   |

**Figures**

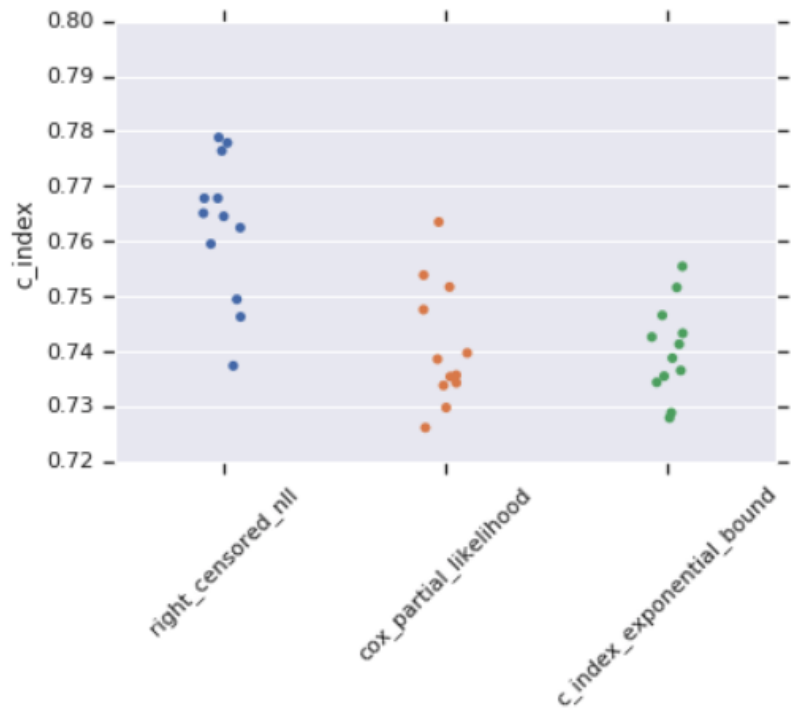

**Supplementary Figure S1. Comparison of loss functions for DLS training.**
We compared three loss functions for DLS training: 1) censored cross-entropy, 2) Cox partial likelihood, 3) exponential lower bound on concordance index with the TCGA KIRC dataset. For each loss function 3 batch sizes (32, 64, 128) and 4 learning rates (10e-3, 5e-4, 10e-4, 5e-5, 10e-5 ) were tried. Models were evaluated on the tune split.